\documentclass[
12pt,
preprint,preprintnumbers,nofootinbib,
groupedaddress,superscriptaddress,amsmath,amssymb]{revtex4}
\usepackage{graphicx}
\usepackage{dcolumn}
\usepackage{bm}
\usepackage{amssymb}
\usepackage{amsmath}
\usepackage{epsfig}    
\usepackage{color}
\usepackage{slashed}
\usepackage{hhline}

\def\be{\begin{equation}}
\def\ee{\end{equation}}
\newcommand{\bea}{\begin{eqnarray}}
\newcommand{\eea}{\end{eqnarray}}
\newcommand{\nn}{\nonumber}

\numberwithin{equation}{section}

\begin{document}

{\begin{flushright}{KIAS-P17066}
\end{flushright}}

\title{Neutrinophilic two Higgs doublet model with dark matter \\
under an alternative $U(1)_{B-L}$ gauge symmetry}
%

\author{Takaaki Nomura}
\email{nomura@kias.re.kr}
\affiliation{School of Physics, KIAS, Seoul 02455, Korea}

\author{Hiroshi Okada}
\email{macokada3hiroshi@cts.nthu.edu.tw}
\affiliation{Physics Division, National Center for Theoretical Sciences, Hsinchu, Taiwan 300}

\date{\today}

\begin{abstract}
We propose Dirac type active neutrino with rank two mass matrix and a Majorana fermion dark matter candidate  with an alternative local $U(1)_{B-L}$ extension of neutrinophilic two Higgs doublet model. Our dark matter candidate can be stabilized due to charge assignment under the gauge symmetry without imposing extra discrete $Z_2$ symmetry and the relic density is obtained from $Z'$ boson exchanging process. Taking into account collider constraints on $Z'$ boson mass and coupling, we estimate the relic density.
\end{abstract}
\maketitle
\newpage

\section{Introduction}

An origin of tiny neutrino masses is one of the unsolved issues in the standard model (SM) where neutrino mass type can be either Dirac or Majorana type.
Majorana type neutrino mass can be realized in many scenarios such as type-I seesaw mechanism in which heavy SM singlet right-handed neutrinos are introduced.
On the other hand, Dirac type neutrino mass can also be obtained as charged leptons by introducing right-handed neutrinos without Majorana mass term.
In such a scenario, neutrinophilic two Higgs doublet model is suggested to avoid large hierarchy in the Yukawa couplings where only one Higgs doublet with small vacuum expectation value (VEV) has Yukawa interaction among lepton doublets and right-handed neutrinos giving neutrino masses~\cite{Davidson:2009ha,Wang:2006jy,Baek:2016wml}. This kind of Higgs doublet model can be constructed by imposing symmetry such as global $U(1)$ symmetry~\cite{Davidson:2009ha,Baek:2016wml}. It is also interesting to consider realization of neutrinophilic two Higgs doublet model based on an exotic $U(1)$ gauge symmetry such as $U(1)_{B-L}$.
Then we consider alternative $U(1)_{B-L}$ charge assignment for right-handed neutrinos~\cite{Montero:2007cd, Ma:2014qra, Singirala:2017see, Nomura:2017vzp} since original $U(1)_{B-L}$ charge assignment is not suitable due to universal $B-L$ charge for leptons including right-handed neutrinos. 

In this paper, we construct a neutrinophilic two Higgs doublet model based on an alternative $U(1)_{B-L}$ gauge symmetry which introduces three right-handed neutrinos $\nu_{R_1}$, $\nu_{R_2}$ and $\nu_{R_3}$ with $B-L$ charge $-4$, $-4$ and $5$ to cancel gauge anomalies.
We also assign $B-L$ charge $-3$ to one of the Higgs doublets, and the $B-L$ charged Higgs doublet only has Yukawa couplings among SM lepton doublets and right handed neutrinos $\nu_{R_{1,2}}$. Thus a Dirac type neutrino mass matrix is obtained after electroweak symmetry breaking and the smallness of neutrino mass can be explained by the small VEV of the $B-L$ charged Higgs doublet by choosing parameters in the scalar potential appropriately.
Furthermore $\nu_{R_3}$ is stabilized by an accidental $Z_2$ symmetry due to the charge assignment and can be a good dark matter (DM) candidate.
{Here we emphasize that three fermion contents inducing active neutrino mass and providing DM are required by anomaly cancellation condition, and our charge assignments for the fermions naturally guarantee the stability of DM without inducing further symmetry such as discrete $Z_2$ symmetry. }
Then we discuss phenomenology of the model such as $Z'$ boson at collider experiments and DM physics.
The constraints on $Z'$ mass and $U(1)_{B-L}$ gauge coupling can be obtained by the data of LEP experiment and the current LHC experiments.
Taking into account the constraints, DM relic density is estimated by calculating annihilation process via $Z'$ exchange in the $s$-channel.

This paper is organized as follows.
In Sec.~II, we show our model, 
and formulate the neutral fermion sector, boson sector and lepton sector.
In Sec.~III, we discuss phenomenology of the model such as $Z'$ boson at collider experiments and dark matter physics.
Finally we conclude and discuss in Sec.~IV.


 \begin{widetext}
\begin{center} 
\begin{table}[t]
\begin{tabular}{|c||c|c|c|c||c|c|c|c|}\hline\hline  
Fields & ~$\Phi$~ & ~$H$~ & ~$\varphi_{10}$~  & ~$\varphi_{3}$~ &~$L_{L_a}$~ & ~$e_{R_a}$~ & ~$\nu_{R_i}$~ & ~$X_{R}$~ 
\\\hline 
 $SU(2)_L$ & $\bm{2}$  & $\bm{2}$  & $\bm{1}$  & $\bm{1}$ & $\bm{2}$ & $\bm{1}$  & $\bm{1}$ & $\bm{1}$   \\\hline 
$U(1)_Y$ & $\frac12$ & $\frac12$  & $0$ & $0$ & $-\frac12$  & $-1$ & $0$ & $0$    \\\hline
 $U(1)_{B-L}$ & $-3$ & $0$  & $10$  & $3$  & $-1$  & $-1$   & $-4$  & $5$   \\\hline
\end{tabular}
\caption{Field contents of bosons and fermions
and their charge assignments under $SU(2)_L\times U(1)_Y\times U(1)_{B-L}$, where $a=1-3$ and $i=1,2$ are flavor indices.}
\label{tab:1}
\end{table}
\end{center}
\end{widetext}

\section{ Model setup and particle contents}
In this section, we introduce our model and discuss masses in scalar sector and fermion sector.
First of all, we introduce the gauged $U(1)_{B-L}$ symmetry, introducing non-trivial $B-L$ charge assignments of the right-handed neutrinos; $(-4,-4,5)$ for $(\nu_{R_1},\nu_{R_2}, X_{R})$ where we write third right-handed neutrino by $X_R$.  First of two right-handed neurtinos $(\nu_{R_1},\nu_{R_2})$ construct the Dirac type active neutrinos via Dirac Yukawa term~\cite{Baek:2016wml}. On the other hand the third right-handed neutrino $X_{R}$ becomes a Majorana fermion by itself after the $B-L$ symmetry breaking.
Thus $X_{R}$ can be a good DM candidate which is stabilized by an accidental $Z_2$ symmetry due to the gauge symmetry~\footnote{At non-renormalizable level, we would have dimension 7 operator such as $\bar \nu_{R_{i}}^c X_{R} \varphi_{10}^*(\varphi_3)^3$. We consider such a term is highly suppressed by sufficiently large cut-off scale and does not affect stability of DM and phenomenology.}.
To construct mass terms for the SM fermions appropriately, we then adopt the neutrinophilic two Higgs doublet model, in which a new isospin doublet boson $\Phi$
with nonzero $B-L$ charge is introduced in order to induce the neutrino masses while the other SM fermion masses are generated with the SM-like Higgs doublet $H$ that is the same as SM.
 In addition, we introduce isospin singlet scalars $\{ \varphi_{10}, \varphi_3 \}$ with nonzero $B-L$ charges denoted by subscripts where $\varphi_{10}$ is required to give $X_{R}$ mass and $\varphi_{3}$ is necessary to avoid massless Goldstone boson from Higgs doublet $\Phi$.
All the field contents and their assignments are summarized in table~\ref{tab:1}.
Under these framework, one finds the following renormalizable Lagrangian:
\begin{align}
-{\cal L}_{L}&=
(y_\ell)_{aa}\bar L_{L_a} e_{R_a} H + (y_\nu)_{ai}\bar L_{L_a} \tilde\Phi \nu_{R_i}
+f_X \bar X^c_{R}  X_{R} \varphi_{10}^* + {\rm c.c.},\\
V&= -\mu_H^2 H^\dag H + \mu_\Phi^2 \Phi^\dag \Phi - \mu^2_{10} \varphi_{10}^* \varphi_{10}^{}  -  \mu^2_{3} \varphi_{3}^*\varphi_{3}^{}
-( \mu \Phi^\dagger H \varphi_3^*  + h.c. ) \nonumber \\
&+\lambda_1 (\Phi^\dag \Phi)^2 +\lambda_2 (H^\dag H)^2  +\lambda_{\varphi_{10}^{}} (\varphi_{10}^* \varphi_{10}^{})^2 
+\lambda_{\varphi_{3}} (\varphi_{3}^* \varphi_{3}^{})^2 \nn \\ &
+ \lambda_{3} (H^\dag H)(\Phi^\dag\Phi)
+\lambda_{4} (H^\dag \Phi)(\Phi^\dag H) 
+ \lambda_{H\varphi_{10}} (H^\dag H)(\varphi_{10}^*\varphi_{10}^{}) + \lambda_{\Phi \varphi_{10}} (\Phi^\dag \Phi)(\varphi_{10}^*\varphi_{10}^{}) \nonumber \\
&
+ \lambda_{H\varphi_{3}} (H^\dag H)(\varphi_{3}^*\varphi_{3}^{}) + \lambda_{\Phi \varphi_{3}} (\Phi^\dag \Phi)(\varphi_{3}^*\varphi_{3}^{})  + \lambda_{\varphi_{3} \varphi_{10}} (\varphi_{3}^*\varphi_{3}^{})(\varphi_{10}^*\varphi_{10}^{}) 
\label{eq:lag-lep}
\end{align}
where we assume the parameters in the potential are real, $\tilde H \equiv (i \sigma_2) H^*$ with $\sigma_2$ being the second Pauli matrix, $a$ runs over $1$ to $3$, and $i$ runs over $1$ to $2$.

\subsection{Scalar sector}
The scalar fields are parameterized as 
\begin{align}
&H =\left[\begin{array}{c}
w^+\\
\frac{v_H + h +i z}{\sqrt2}
\end{array}\right],\quad 
\Phi =\left[\begin{array}{c}
\phi^+\\
\frac{v_\phi + \phi_R + i \phi_I}{\sqrt2}
\end{array}\right],\quad 
\varphi_{Q}=
\frac{v_{Q} + \varphi_{QR} + iz'_Q}{\sqrt2},
\label{component}
\end{align}
where $Q = \{3, 10 \}$ indicates $B-L$ charges for singlet scalar fields, the lightest mass eigenstate after diagonalizing the matrix in basis of $(w^\pm$ , $\phi^\pm)$, which is massless, is absorbed by the SM $W^\pm$ boson, and two of the mass eigenstate in CP odd boson sector $(\phi_I, z, z'_{Q})$ are also absorbed by the SM $Z$ boson and $B-L$ gauge boson $Z'$, as Nambu-Goldstone(NG) bosons.
The VEVs are obtained by applying the conditions $\partial V/ \partial v_Q =0$, $\partial V/ \partial v_H =0$ and $\partial V/ \partial v_\phi =0$ such that 
\begin{align}
&v_{10} \simeq \sqrt{\frac{\mu_{10}^2}{\lambda_{\varphi_{10}}}}, \quad 
v_3 \simeq \sqrt{\frac{2\mu_3^2  - \lambda_{H \varphi_{3}} v_{H}^2 - \lambda_{\varphi_3 \varphi_{10}} v_{10}^2 }{2\lambda_{\varphi_3}}}, \quad 
v_H \simeq \sqrt{\frac{2\mu_H^2  - \lambda_{H \varphi_{3}} v_{3}^2 - \lambda_{H \varphi_{10}} v_{10}^2 }{2\lambda_H}} \nonumber \\
&  v_\phi \simeq \frac{\sqrt{2} \mu v_H v_3 }{2 \mu_\Phi^2 + (\lambda_{3} + \lambda_{4})v_H^2 + \lambda_{\Phi \varphi_{3}} v_{3}^2+ \lambda_{\Phi \varphi_{10}} v_{10}^2 }
\end{align}  
where we assume the relation $ v_{\phi}^2 \ll \{v_{H}^2, v_3^2, \mu_3^2 \} \ll \{v_{10}^2,  \mu_{10}^2\}$. The small $v_\phi$ can be realized taking trilinear coupling $\mu$ to be sufficiently small.
Note that $z'_{10}$ is dominant component of NG boson which is absorbed by $Z'$ boson since we consider VEV of $\varphi_{10}$ is much larger than the others.
After $\varphi_{10}$ developing VEV at high energy scale, our scalar sector has the same structure as discussed in ref.~\cite{Baek:2016wml}.
Then CP-odd component of $\varphi_3$; $z'_{3}$, becomes physical massless Goldstone boson(GB) due to a global symmetry in the scalar potential.
However the existence of this physical Goldstone boson does not cause serious problem in particle physics or cosmology since it does not couple to SM particles directly except for Higgs boson whose couplings are well controlled by the parameters in the potential, and decouples from thermal bath in early Universe.
{Here we discuss the condition that GB decouples from thermal bass at sufficiently early Universe following discussion in ref.~\cite{Weinberg:2013kea}. Since scalar and gauge bosons which couples to GB are heavy their interactions with GB decouple at sufficiently high temperature. We thus focus on interaction between GB and the SM fermions. The ratio between collision and expansion rates is roughly given by $R(T) \sim \lambda_{\varphi_{3} H}^2 m_f^2 T^5 m_{pl}/(m_{\varphi_{3R}}^4 m_h^4 )$~\cite{Weinberg:2013kea} where  we take Boltzmann constant as 1, $m_{pl}$ is Planck mass, and $m_f$ denotes an SM fermion mass; the process GB GB $\leftrightarrow ff$ is induced by the Higgs-$\varphi_3$ mixing for $m_f < T$. The decoupling occurs when $R(T) \sim 1$ and we obtain decoupling temperature $T_d \sim  0.42/\lambda_{\varphi_{3} H}^{2/5}$ GeV assuming $m_f = m_\tau$. In such a case $\lambda_{\varphi_3 H} = 0.001$ gives $T_d \sim 2.7$ GeV which is consistent with condition $T > m_\tau$. Thus if $\lambda_{\varphi_3 H} \lesssim 0.001$ GB decouples from thermal bath around $O(1)$ GeV temperature and cosmology is not affected by GB; note that SM fermions with $m_f < m_\tau$ decouples earlier due to smaller couplings.}

In our scenario, one finds that $v\equiv \sqrt{v_H^2+v_\phi^2}\sim v_H$ where $v \simeq 246$ GeV since $v_\phi$ is expected to be tiny in order to generate the active neutrino masses.
Thus charged component $w^\pm$ in $H$ is approximated to be NG boson which is absorbed by $W^\pm$ boson while the $\phi^\pm$ from $\Phi$ is physical charged Higgs boson.
Similarly the CP-odd boson $z$ is absorbed by the neutral SM gauge boson $Z$ while $\phi_I$ is physical CP-odd Higgs. 
The masses of physical charged Higgs and CP-odd Higgs are approximately given by~\cite{Baek:2016wml}
\begin{align}
m_{\phi^\pm}^2 & \simeq \frac{v_2 (\sqrt{2} \mu v_3 - \lambda_4 v_1 v_2)}{2 v_1}, \\
m_{\phi_I}^2 & \simeq \frac{\mu v_2 v_3}{\sqrt{2} v_1}.
\end{align}
The mass matrix for the CP-even scalars has $4 \times 4$ structure in a basis of $(h, \phi_R, \varphi_{3R}, \varphi_{10R})$.
In our analysis, we omit details of the matrix assuming SM Higgs is the lightest component among four physical CP-even scalar bosons.
 In addition, we assume mixing among SM Higgs and other CP-even scalars are small to avoid experimental constraints for simplicity.
More details of the scalar sector can be found in refs.~\cite{Davidson:2009ha,Baek:2016wml}, and we focus on $Z'$ and DM physics in the following analysis below.

\subsection{Fermion sector}
The masses for charged-leptons are induced via  $y_\ell$ after symmetry breaking,
and active neutrino masses are also done via $y_\nu$  term where neutrinos are supposed to be Dirac type fermions.
Their masses are symbolized by $m_{\ell_a}\equiv v_H y_{\ell_a}/\sqrt2$ and $m_{\nu_{ai}}\equiv v_\phi y_{\nu_{ai}}/\sqrt2$.
Since the charged-lepton mass matrix is diagonal, the neutrino mixing matrix $V_{ab}$ is arisen from diagonalizing the neutrino mass matrix squared as 
\begin{equation}
(M_\nu^{\rm diag})^2 = (V^\dagger)_{aa'} \sum_{i=1-2}(m_{\nu_{a' i}}m_{\nu_{ib'}}^\dag) V_{b' b}, 
\end{equation}
where $V$ is measured PMNS matrix by the neutrino oscillation data~\cite{Gonzalez-Garcia:2014bfa}.
Notice here that one of the active neutrinos is massless due to the rank two matrix.
{Thus one can parametrize the neutrino mass matrix in terms of observables and arbitrary parameters in the following form:
\begin{equation}
m_{\nu_{a'i}} = V_{aa'} (M_\nu^{\rm diag})_{a'a'} {\cal O}_{a' i} , 
\end{equation}
where ${\cal O}_{a' i}$ is generally an arbitrary three by two matrix with complex values, satisfying ${\cal O}{\cal O}^\dag\neq1_{3\times3}$ and ${\cal O}^\dag{\cal O}=1_{2\times2}$.
However since we have enough theoretical parameters,  
we simply reduce the parameterization of ${\cal O}$ for the normal hierarchy (NH) and inverted hierarchy (IH)~\cite{Rink:2016vvl}, which is analogy of the case of Majorana neutrino mass matrix:
\begin{align}
O =\left[\begin{array}{cc}
0 & 0\\
\cos z & -\sin z \\
\zeta \sin z & \zeta\cos z \\
\end{array}\right], \quad 
O =\left[\begin{array}{cc}
\cos z & -\sin z \\
\zeta \sin z & \zeta \cos z \\
0 & 0\\
\end{array}\right],  
\label{eq:omix}
\end{align}  
respectively,
{where $\zeta$ is complex number satisfying $|\zeta|=1$, and we parametrize $z$ to be real value running over $z \in [0, 2\pi]$ and $\zeta = e^{i \theta}$ with $\theta \in [0, 2\pi]$.}
In our numerical analysis, we will use the global fit of the current neutrino oscillation data as the best fit values for NH and IH~\cite{Gonzalez-Garcia:2014bfa}:
\begin{align}
{\rm NH}:\ 
&s_{12}^2=0.304,\quad s_{23}^2=0.452,\quad s_{13}^2=0.0218,\quad \delta_{CP}=\frac{306}{180}\pi,\nn\\
&\quad (m_{\nu_1},\ m_{\nu_2},\ m_{\nu_3})\approx(0, 8.66, 49.6)\ {\rm meV}, \\ 
{\rm IH}:\ 
&s_{12}^2=0.304,\quad s_{23}^2=0.579,\quad s_{13}^2=0.0219,\quad \delta_{CP}=\frac{254}{180}\pi,\nn\\
&\quad (m_{\nu_1},\ m_{\nu_2},\ m_{\nu_3})\approx(49.5,50.2,0)\ {\rm meV},
\end{align}
where $s_{12,13,23}$ are the short-hand notations of $\sin\theta_{12,13,23}$ for three mixing angles of $V$,
while two Majorana phases are taken to be zero.
We show some samples of allowed regions for $z$ and the mass scale of $m_{\nu}$ in Fig.~\ref{fig:neut-osci},
where upper figures represent NH and lower one do IH, while the left figures shows the mass scale of $m_{\nu_{11}}$ in terms of $z$ and the right ones do the mass scale of $m_{\nu_{11}}$ and $m_{\nu_{22}}$. 
{They suggest that typical mass scale of $m_\nu$ is $10^{-12} \sim 10^{-11}$ GeV~\footnote{We found that the other mass parameters of $m_\nu$ are also of the order $10^{-12} \sim 10^{-11}$ GeV.}. Note here that correlations between them seem to occur due to the manner of our parametrization. 
Since $(m_\nu)_{ai} = v_\phi y_{\nu_{ai}}/\sqrt{2}$, the order of Yukawa coupling is $10^{-12\sim11}/v_\phi$; when $v_\phi \sim \mathcal{O}({\rm KeV})$ the order of coupling is around $10^{-6}$ which is similar to electron Yukawa coupling. }

\begin{figure}[t]
\begin{center}
\includegraphics[width=80mm]{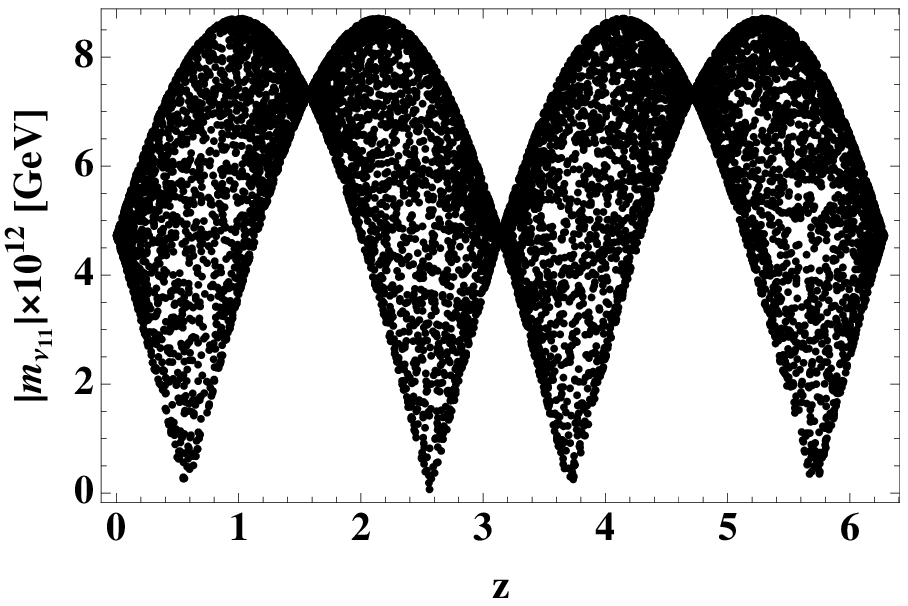} 
\includegraphics[width=80mm]{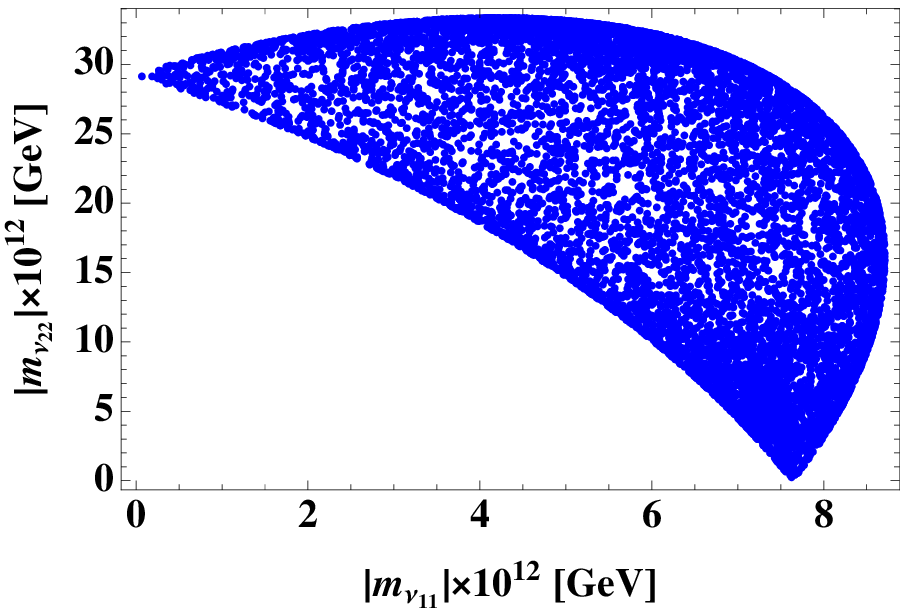} \\
\includegraphics[width=80mm]{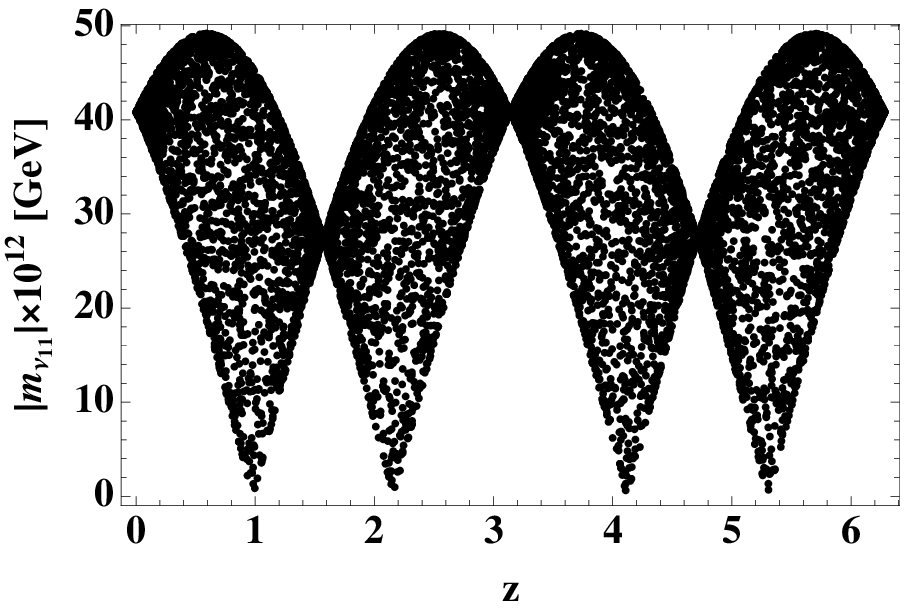} 
\includegraphics[width=80mm]{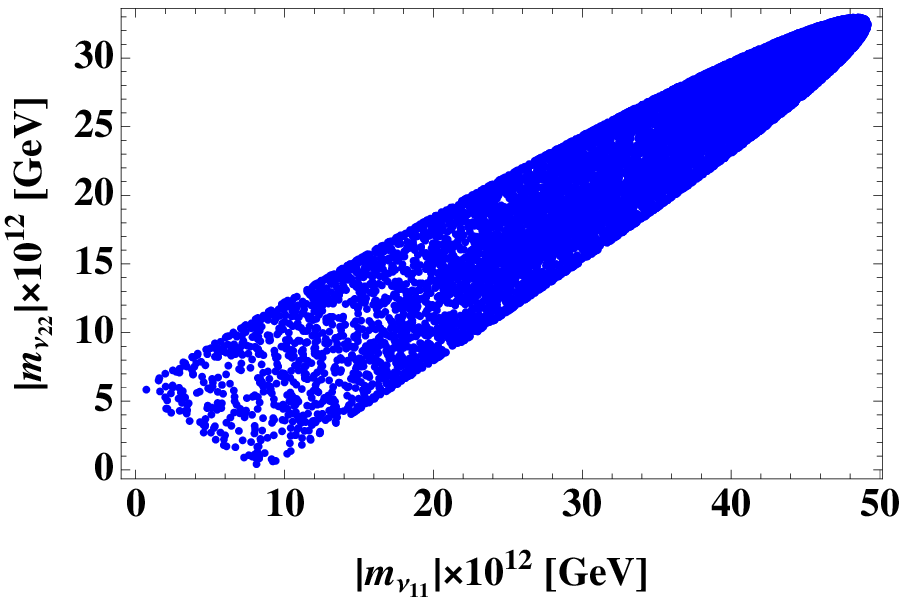}
\caption{The allowed regions to satisfy the neutrino oscillation data, where upper figures represent NH and lower one do IH, while the left figures shows the mass scale of $m_{\nu_{11}}$ in terms of $z$ and the right ones do the mass scale of $m_{\nu_{11}}$ and $m_{\nu_{22}}$.}   \label{fig:neut-osci}
\end{center}\end{figure}

}

The third right-handed neutrino obtains Majorana mass term from the term with $f_X$ after $\varphi_{10}$ developing the VEV: $\langle \varphi_{10} \rangle = v_{10}/\sqrt{2}$.
The Majorana mass is simply given by 
\begin{equation}
M_X = \frac{f_X}{\sqrt{2}} v_{10}.
\end{equation}

\section{Phenomenology}

In this section, we consider phenomenology of our model focusing on $Z'$ boson and dark matter candidate.

\subsection{$Z'$ boson}

Here we consider constraints from collider experiments for $Z'$ boson mass $m_{Z'}$ and $U(1)_{B-L}$ gauge coupling $g_{BL}$.
The gauge interactions of $Z'$ and fermions are given by
\begin{align}
-{\cal L}&=
5g_{BL} \bar X\gamma^\mu P_R X Z'_\mu + g_{BL} Q^f_{BL} \bar f_{SM} \gamma^\mu f_{SM} Z'_\mu\nn\\
& - g_{BL} (\bar\nu_a \gamma^\mu P_L\nu_a + 4 \bar\nu_i \gamma^\mu P_R\nu_i) Z'_\mu,
\label{eq:dmint}
\end{align}
where $Q^f_{BL}$ is the charge of $U(1)_{B-L}$ symmetry,  and $f_{SM}$ denotes all the electrically charged fermions in SM.
Here $Z'$ mass is given by $m_{Z'} \simeq 10 g_{BL} v_{10}$ as we assume $v_{10} \gg \{v_{3}, v_{\phi} \}$.

Firstly, we discuss constraint from the LEP experiment.
Since $Z'$ couples to SM leptons, we obtain the following effective interactions considering $Z'$ is sufficiently heavy;
\begin{equation}
L_{eff} = \frac{1}{1+\delta_{e \ell'}} \frac{g_{BL}^2}{m_{Z'}^2}  (\bar e \gamma^\mu e)( \bar \ell' \gamma_\mu \ell')
\end{equation}
where $\ell' = e$, $\mu$ and $\tau$.
In this case, we obtain constraints from the analysis for the process $e^+ e^- \to \ell'^+ \ell'^-$ with the data of measurement at LEP~\cite{Schael:2013ita}: 
\begin{align}
\frac{m_{Z'}}{g_{BL}}   \gtrsim  6.9\ {\rm TeV}.\label{eq:lep}
\end{align}
In our following analysis, we take into account the constraint. 

\begin{figure}[t]
\begin{center}
\includegraphics[width=80mm]{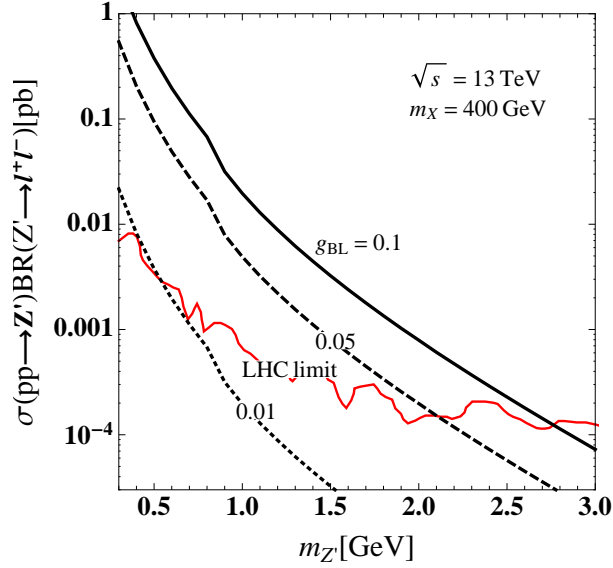} 
\caption{$\sigma(pp \to Z')BR(Z' \to \ell^+ \ell^-)$, $\ell = e$ and $\mu$, as a function of $m_{Z'}$ at $\sqrt{s} = 13$ TeV with $g_{BL}=0.1$, $0.05$ and $0.01$ as indicated in the plot. The red curve shows the upper limit from the ATLAS experiment~\cite{Aaboud:2017buh}.}   \label{fig:ZpLHC}
\end{center}\end{figure}

We next discuss the $Z'$ production at LHC.
In hadron collider experiments, $Z'$ boson can be produced via the process $q \bar q \to Z'$ where $q$ indicates SM quarks. 
Here we estimate the production cross section using CalcHEP~\cite{Belyaev:2012qa} implementing the relevant interactions with the CTEQ6 parton distribution functions (PDFs)~\cite{Nadolsky:2008zw}. Then $Z'$ decays into $B-L$ charged particles where we only consider fermions assuming masses of scalar bosons are greater than $m_{Z'}/2$.
The decay width is given by
\begin{align}
\Gamma_{Z'} &= \frac{g_{BL}^2m_{Z'}}{12 \pi} \sum_{f} N_c^{f} C_f |Q_{BL}^f|^2 \left( 1 + \frac{2 m_f^2}{m_{Z'}^2} \right) \sqrt{1- \frac{4 m_f^2}{m_{Z'}^2}} \nonumber \\
& \simeq \frac{g_{BL}^2 m_{Z'}}{12 \pi} \left[ \frac{133}{6} + \frac{1}{3} \left( 1 + \frac{2 m_t^2}{m_{Z'}^2} \right) \sqrt{1- \frac{4 m_t^2}{m_{Z'}^2}}
+ \frac{25}{2} \left( 1 + \frac{2 m_X^2}{m_{Z'}^2} \right) \sqrt{1- \frac{4 m_X^2}{m_{Z'}^2}} \right],
\label{eq:ZpWidth}
\end{align}
where $f$ denotes any fermion in the model, $N_c^f$ is color factor, and we used $m_{f}/m_{Z'} \ll 1$ for fermions except for top quark and $X$.
The branching ratio for a mode $Z' \to f \bar f$ is given by
\begin{equation}
BR(Z' \to f \bar f) \simeq \frac{g_{BL}^2m_{Z'}}{12 \pi} N_c^{f} C_f |Q_{BL}^f|^2 \left( 1 + \frac{2 m_f^2}{m_{Z'}^2} \right) \sqrt{1- \frac{4 m_f^2}{m_{Z'}^2}} \times \Gamma_{Z'}^{-1}.
\end{equation}
The branching ratio for charged lepton modes are less than $\sim 0.05$ since $Z'$ dominantly decays into $\nu_{R_{1,2}}$ due to the charge assignment.
In Fig.~\ref{fig:ZpLHC}, we show the product of cross section and branching ratio, $\sigma(pp \to Z'){\rm BR}(Z' \to \ell^+ \ell^-)$ where $\ell = e$ and  $\mu$, at $\sqrt{s} = 13$ TeV as 
a function of $m_{Z'}$ with $g_{BL} = \{0.1, 0.05, 0.01\}$ which is compared with the limit from the LHC data~\cite{Aaboud:2017buh}.
We thus find that $m_{Z'} \gtrsim 2.8$ TeV is required for $g_{BL} = 0.1$ while $m_{Z'} \simeq 500$ GeV is still allowed for smaller gauge coupling $g_{BL} = 0.01$.
Note that the constraint on $Z'$ mass is weaker than that in original $U(1)_{B-L}$ case~\cite{Klasen:2016qux} since our $Z'$ dominantly decays into light right-handed neutrinos $\nu_i$ which has larger charge than the other SM fermions. More parameter region will be tested by the data of the future LHC experiments with larger integrated luminosity.

\subsection{ Dark matter} 
In this subsection we discuss a dark matter candidate; $X_R$.
{\it Firstly, we assume that any contributions from the Higgs mediating interactions are negligibly small and DM annihilation processes are dominated by the gauge interaction with $Z'$; we thus can easily avoid the constraints from direct detection searches such as LUX~\cite{Akerib:2016vxi}, XENON1T~\cite{Aprile:2017iyp}, and PandaX-II~\cite{Fu:2016ega, PandaX}.} 
{Here we discuss the condition for Higgs portal interaction from direct detection constraints. The nucleon-DM interaction is induced by Higgs portal interaction via mixing between the SM Higgs and $\varphi_{10}$. The relevant effective interaction is given by~\cite{Baek:2016wml}
\begin{equation}
L_{\rm eff} = \sum_q \frac{f_X m_q \sin \theta \cos \theta}{2 \sqrt{2} v m_h^2} \bar X X \bar q q, 
\end{equation}
where $\theta$ is mixing angle for Higgs-$\varphi_{10}$ mixing, $m_q$ is a mass of quark $q$ and we assumed heavier scalar boson is much heavier than $m_h$. Then $X$-$N$ spin independent scattering cross section is obtained as 
\begin{equation}
\sigma_{XN} \simeq \frac{1}{2 \pi} \frac{\mu_{NX}^2 f_N^2 m_N^2 f_X^2 \sin^2 \theta \cos^2 \theta}{v^2 m_h^4}
\end{equation}
where $m_N$ is the nucleon mass, $\mu_{NX} = m_N m_X/(m_N+m_X)$ is the reduced mass of nucleon and DM and $f_N$ is effective coupling constant for Higgs-Nucleon interaction. 
For simplicity, we estimate the cross section applying $f_n \simeq 0.287$ for neutron. Finally we estimate the cross section as 
\begin{equation}
\sigma_{Xn} \sim 4.2 \times 10^{-43} \left( \frac{100 \, {\rm GeV}}{m_X} \right)^2 f_X^2 \sin^2 \theta \cos^2 \theta \  {\rm cm}^2.
\end{equation}
Therefore direct detection constraints can be satisfied with small mixing angle such as $\sin \theta \ll 0.1$ even if coupling $f_X$ is $\mathcal{O}(1)$.}
{We next consider direct detection via $Z'$ exchange.
The relevant effective interactions between DM and the SM quarks is given by
\begin{equation}
\label{eq:DMqq}
\frac{g_{BL}^2}{m_{Z'}^2} \frac{5}{6} (\bar X \gamma^\mu \gamma_5 X)(\bar q \gamma^\mu q) \equiv \frac{(\bar X \gamma^\mu \gamma_5 X)(\bar q \gamma^\mu q)}{\Lambda_{Z'}^2}, 
\end{equation}
where DM has only axial vector current due to Majorana property and we defined $\Lambda_{Z'} \equiv 6 m_{Z'}/(5 g_{BL})$.
The operator in Eq.~(\ref{eq:DMqq}) induces spin dependent operator $\vec{s}^\bot_X \cdot \vec{q}$ and $\vec{s}_X \cdot (\vec{s}_N \times q)$~\cite{Liu:2017kmx,Anand:2013yka,Fan:2010gt}; 
$\vec{q}$ is transferred momentum, $s_{X(N)}$ is spin operator of DM(nucleon) and $\bot$ indicate direction perpendicular to $\vec{q}$ direction.
In Ref.~\cite{Liu:2017kmx}, the lower limit of $\Lambda_{Z'}$ is given as $\sim 1$ TeV which is obtained by data from PandaX, LUX and XENON1T including spin-dependent direct detection results~\cite{Fu:2016ega}. 
This constraint is much weaker than the constraint from collider search of $Z'$ as shown in Fig.~\ref{fig:ZpLHC}. Thus constraint from direct detection is not stringent in our model.  }

{\it Relic density}: We have annihilation modes via gauge interaction as $X \bar X \to Z' \to f \bar f$ to explain the relic density of DM:
$\Omega h^2\approx 0.12$~\cite{Ade:2013zuv}, and their relevant Lagrangian in basis of mass eigenstate is given in Eq.~(\ref{eq:dmint}). 
Then the relic density of DM is estimated by~\cite{Edsjo:1997bg}
\begin{align}
&\Omega h^2
\approx 
\frac{1.07\times10^9}{\sqrt{g_*(x_f)}M_{Pl} J(x_f)[{\rm GeV}]},
\label{eq:relic-deff}
\end{align}
where $g^*(x_f\approx25)$ is the degrees of freedom for relativistic particles at temperature $T_f = M_X/x_f$, $M_{Pl}\approx 1.22\times 10^{19}$ GeV,
and $J(x_f) (\equiv \int_{x_f}^\infty dx \frac{\langle \sigma v_{\rm rel}\rangle}{x^2})$ is given by~\cite{Nishiwaki:2015iqa}
\begin{align}
J(x_f)&=\int_{x_f}^\infty dx\left[ \frac{\int_{4M_X^2}^\infty ds\sqrt{s-4 M_X^2} W(s) K_1\left(\frac{\sqrt{s}}{M_X} x\right)}{16  M_X^5 x [K_2(x)]^2}\right],\\ 
W(s)
\approx &\frac{4}{3\pi} (s-M_X^2) \left| \frac{5 g_{BL}^2 }{s-m_{Z'}^2+i m_{Z'} \Gamma_{Z'}}\right|^2
\sum_f  C_f\sqrt{1-\frac{4 m_{f}^2}{s}} (s+2 m^2_{f})|Q_{BL}^{f}|^2
\label{eq:relic-deff}
\end{align}
where we implicitly impose the kinematical constraint above, $C_f = 1/2$ for neutrino pairs including two light right-handed neutrino $\nu_i$ in the second line of Eq.~(\ref{eq:dmint}) otherwise $C_f =1$, and the width of $Z'$ is given by Eq.~(\ref{eq:ZpWidth}).

\begin{figure}[t]
\centering
\includegraphics[width=10cm]{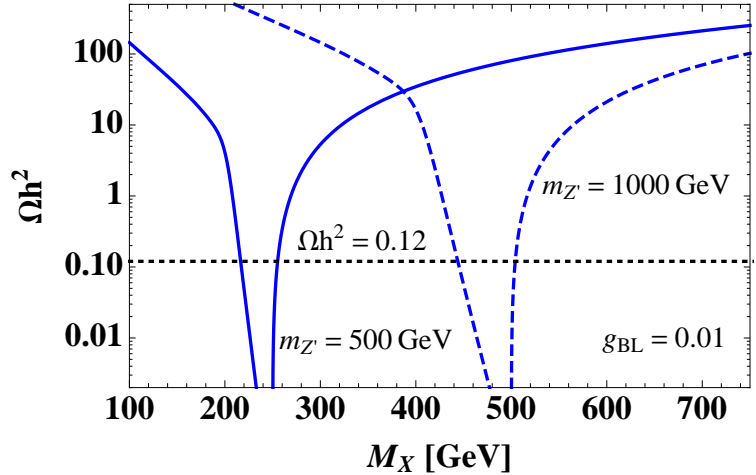}
\caption{The relic density of dark matter $X$ as a function of $m_{X}$ where solid(dashed) lines correspond to $m_{Z'} = 500(1000)$ GeV and $g_{BL} = 0.01$ is applied as the allowed sampling point.  }
\label{fig:relic}
\end{figure}

In fig.~\ref{fig:relic}, we show the relic density in terms of $M_X$, where we fix parameters $m_{Z'}=\{500, 1000 \}$ GeV and $g_{BL}=0.01$ which are allowed by the collider experiments.
We find that the correct relic density can be obtained near the $Z'$ pole since we need resonant enhancement due to small gauge coupling.

\section{Conclusion}
In this paper, we have discussed neutrinophilic two Higgs doublet model with alternative anomaly free $U(1)_{B-L}$ gauge symmetry under which three right-handed neutrinos $\nu_{R_1}$, $\nu_{R_2}$ and $\nu_{R_3}$ have charges $Q_{B-L} = -4$, $-4$ and $5$. The neutrnophilc structure is realized by assigning non-zero $U(1)_{B-L}$ charge to one of the Higgs doublets due to the charge assignment for the right-handed neutrinos. Then two right-handed neutrinos $\nu_{R_{1,2}}$ have Yukawa coupling with SM left-handed neutrinos and we obtain $2 \times 3$ Dirac neutrino mass matrix predicting one massless neutrino. In addition, $X_R \equiv \nu_{R_3}$ can be a good dark matter candidate since it is stabilized by an accidental $Z_2$ symmetry in our model due to the charge assignment of $U(1)_{B-L}$. In the scalar sector, we have introduced two SM singlet scaler fields $\varphi_{10}$ and $\varphi_3$ with $U(1)_{B-L}$ charge $10$ and $3$, respectively, where the former one is introduced to break $U(1)_{B-L}$ giving $Z'$ boson mass and the latter one is introduced to avoid massless Goldstone boson from Higgs doublet. Then CP-odd component of $\varphi_3$ becomes physical Goldstone boson which is harmless since it does not directly couples to SM particles except for Higgs boson and the coupling to Higgs boson can be controlled by the parameters in the scalar potential.

 Then we have discussed phenomenology of the model such as $Z'$ boson production at collider experiments and dark matter physics.
 Our $Z'$ can be produced at LHC and can decay into SM leptons. We thus have estimated the cross section and branching ratios for $Z' \to \ell^+ \ell^-$ and compared resulting values of the product of the cross section and branching ratio with the current LHC constraint. We find that $Z'$ should be heavier than $\sim 2.8$ TeV for gauge coupling $g_{BL} = 0.1$ while $m_{Z'} \simeq 500$ GeV is still allowed for smaller gauge coupling $g_{BL} = 0.01$.
We have found that the constraint on $Z'$ mass is weaker than that in original $U(1)_{B-L}$ case,
since our $Z'$ dominantly decays into light right-handed neutrinos $\nu_i$ that has larger charge than the other SM fermions.
We have also estimated relic density of dark matter which is determined by the thermally averaged cross section of the processes $XX \to Z' \to f f$ where $f$ is any fermions in the model. We have shown that the observed relic density can be obtained with $Z'$ mass and $g_{BL}$ allowed by the collider constraints.
 Our model can be further tested in future by both collider and dark matter
 experiments. 

\section*{Acknowledgments}
\vspace{0.5cm}
H. O. is sincerely grateful for the KIAS member and all around.

\end{document}